\documentclass{IPEC2026TEC}
\IEEEoverridecommandlockouts
\usepackage{cite}
\usepackage{amsmath,amssymb,amsfonts}
\usepackage{algorithmic}
\usepackage[dvipdfmx]{graphicx}
\usepackage{textcomp}
\usepackage{xcolor} % If you are using uplatex or similar tools, please add the [dvipdfmx] option.
\def\BibTeX{{\rm B\kern-.05em{\sc i\kern-.025em b}\kern-.08em
    T\kern-.1667em\lower.7ex\hbox{E}\kern-.125emX}}
\begin{document}

\title{Unlocking Embodied Probabilistic Computational Features in Motor Drives}

\author{\IEEEauthorblockN{
    Subham Sahoo\textsuperscript{1}*, 
    Huai Wang\textsuperscript{1}, 
    Frede Blaabjerg\textsuperscript{1}
}
\IEEEauthorblockA{1 Department of Energy, Aalborg University, Denmark}
\IEEEauthorblockA{*email: sssa@energy.aau.dk}
}

\maketitle

\begin{abstract}
Artificial intelligence (AI)-driven fault diagnosis in motor drives often requires significant computational efforts and time for re-training, in addition to the limited knowledge behind the model and suitability of training and learning mechanisms. This work bridges this gap by proposing a
structured mechanism of transforming untapped labeled fault data into AI parameters to
leverage probabilistic data-driven learning. This novel \textit{AI reservoir modeling} framework for power electronics not only eliminates exogenous efforts behind learning data patterns and its optimization, but also provides intuitive guidelines for power electronics engineers behind sizing of AI models. This alignment between data and system physics makes the proposed model transparent and interpretable, bridging practical understanding with data-driven learning. Its computational efficiency is demonstrated using experimental data that structured, physics-aware reservoirs achieve higher diagnostic accuracy and clearer explanations than conventional black-box AI methods.
\end{abstract}


\section{Introduction}
Model based fault detection in motor drives can be an intractable challenge, since heterogeneous degradation of components leads to inaccurate residuals for fault indicator \cite{yliao,roja}. Hence, artificial intelligence (AI) methods--primarily deep learning (CNNs, LSTMs, Transformers) and hybrid approaches that transform vibration/current/time–frequency signals into image-like representations for end-to-end classification \cite{b2,b3} are increasingly used to provide a structured assessment for classification of faults using historic data \cite{bnn1,b4}. However, the lack of scientific knowhow behind suitability of AI training and learning mechanisms, as shown in Fig. 1(a), for different features in the power electronics (PE) dataset is still \textit{limited}. Furthermore, their sensitivity to changing load/speed conditions require multiple sensors and heavy pre-processing, exhibiting limited interpretability \cite{bnn1}. Hence, unstructured data-driven exploration of fault/degradation signatures in PE leads to:

\begin{enumerate}
    \item \textit{\textbf{Over-confident results:}} Over-designed AI models may lead to misinterpret/overfitting correlations as causation without proper domain PE knowledge constraints. This usually results in over-confident predictions without any traces of uncertainties.
    \item \textit{\textbf{High computations:}} Regardless of the dataset size, {over-designing of AI models} often lead to large-scale models, resulting in high computational time and resources \cite{bnn1}.
\end{enumerate}
\begin{figure}[t]
   \centering
    \includegraphics[scale=2.1]{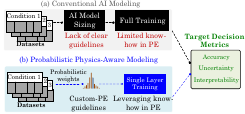}
   \caption{Proposed physics-aware reservoir modeling to bridge the scientific know-how between power electronics (PE) and AI modeling--this circumvents the unclear model sizing and training guidelines for PE applications.}
    \label{Fig_2}
\end{figure}
From a PE perspective, each operating condition of a motor drive—defined by load, speed, and fault state—produces characteristic patterns in measurable signals such as stator currents, torque, and DC-link voltage. Rather than treating these signals as abstract data points, they can be interpreted as realizations of an underlying dynamical system. Over time, these realizations form statistical distributions that capture the intrinsic variability of the system under each fault condition.

To bridge this design gap, this paper proposes a novel mechanism that transforms the motor drive fault data into probabilistic weights and biases collected under different loading conditions, as shown in Fig. 1(b) -- to exploit them as untapped \textit{AI model parameters}. Since each labeled dataset (fault type, operating condition, etc.) represents a distribution of responses rather than a single point, its empirical histogram can define the probability density for weights connecting features to that fault condition. For instance, a consistent increase in torque variance under a chipped tooth fault directly encodes a physical relationship between mechanical degradation and electromagnetic response. 

By mapping these histograms into probabilistic weights, the proposed approach transforms measured system behavior into model parameters without requiring extensive optimization. In this sense, the motor drive itself acts as an \textit{embodied computational system}, where the dynamics of the physical process define a high-dimensional feature space. The AI model does not learn this space from scratch; instead, it extracts and organizes it into a structured representation. This perspective aligns with \textit{reservoir computing} principles \cite{reserv,reserv1}, where complex system dynamics are leveraged as a computational resource. However, unlike conventional reservoir approaches that rely on randomly initialized networks, the proposed method constructs the reservoir using physics-aware statistical representations derived directly from measured data. As a result:
\begin{itemize}
    \item Each weight then represents a statistical confidence in a particular dynamic relationship (e.g., current–torque coupling) observed under that fault.
\item When used in a Bayesian or reservoir framework, this provides transparent fault sensitivity, because the model’s uncertainty or activation can be traced back to physical histogram features (e.g., skewness, variance under fault).
\end{itemize}

The inherent fault signatures in motor drive datasets are leveraged to converge with \textit{physics-aware training} of a single readout layer for fine tuning of their embodied probabilistic dependencies. In this manner, this tailored design mechanism not only reduces the computational efforts and dependence on data science preliminaries, but also bridges the knowledge gap between the modeling of AI in power electronics.
\begin{figure}[t]
    \centering
    \includegraphics[scale=0.55]{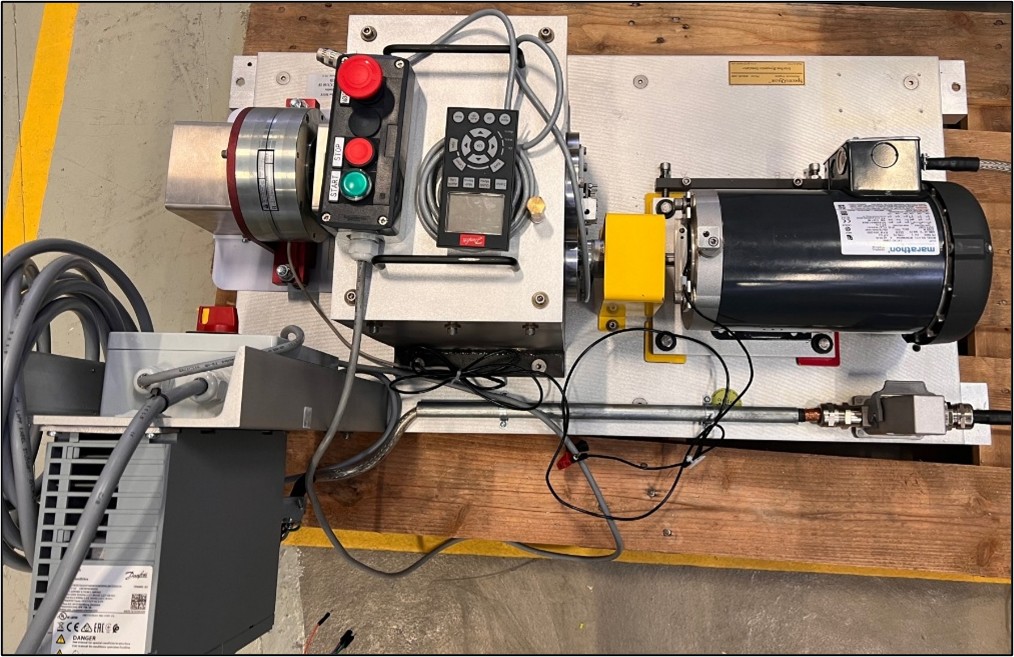}
    \caption{Experimental testbed with SpectraQuest Gearbox Dynamic Simulator and Danfoss VLT Drive FC-103 for collection of data in Table I.}
    \label{Fig_32}
\end{figure}
\begin{figure}[t]
    \centering
    \includegraphics[scale=2.3]{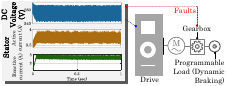}
    \caption{Detailed schematic of Fig. 2 -- Layout of Gearbox Dynamic Simulator \cite{gds} for collection of different gear fault (re-configurable) data, obtained under different loading conditions.}
    \label{Fig_3}
\end{figure}
\section{System Specifications}
SpectraQuest’s Gearbox Dynamics Simulator (GDS), shown in Fig. 1, is used to simulate industrial gearboxes and emulate gear faults. More details on this setup can be found in \cite{agni}.
Gear faults, that typically manifest as cracks on the gear or wear and tear of the gear teeth, were pre-configured for a dynamic brake as programmable load for data collection under different conditions. Hence, the gearbox in Fig. 1 can be easily swapped and reconfigured to the faults in Table I.
\begin{table}[h!]
\centering
\caption{Gear Fault Data Preliminaries}
\begin{tabular}{c|c||c}
\hline
\textbf{Fault Label} & \textbf{Fault} & \textbf{Intrinsic Data} \\
\hline
1 & Missing tooth & Speed  \\
2 & Chipped tooth & Motor Torque \\
3 & Root crack & DC link voltage\\
4  & Surface crack & Active stator current\\
5 & Eccentricity & Reactive stator current\\
\hline
\end{tabular}
\end{table}

Two types of sensors were used in this project, \textit{intrinsic} (within the drive) and \textit{extrinsic} (exterior to the drive). The setup in Fig. 2 and 3 was modified with a Danfoss VLT Drive FC-103 to provide the intrinsic measurements listed in Table I, each sampled at a frequency of 5 kHz. On the other hand, a pair of extrinsic orthogonally aligned analogue accelerometers ADXL1001 are mounted on the gearbox.
\begin{figure*}[t]
    \centering
    \includegraphics[scale=3.4]{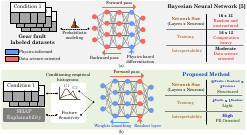}
    \caption{Comparison between (a) conventional physics-informed Bayesian Neural Networks (BNNs) \cite{bnn1}, and (b) the proposed reservoir-inspired approach for fault diagnosis in motor drives--the network size of the proposed method is opportunistically reduced to the number of fault classes.}
    \label{Fig_3}
\end{figure*}
\section{Integrating AI and Physics in PE}
Recent advances in physics-informed AI \cite{pinn1,pinn2} have demonstrated the potential of combining domain physics in PE with data-driven modeling to improve prediction accuracy and interpretability. Such fusion allows learning architectures to capture nonlinear relations in vibration, current, and torque signatures while constraining the solution space through governing physics. In the \textit{Bayesian Neural Network (BNN)} shown in Fig. 4(a), probabilistic modeling of weights:
\begin{equation}
    w_i \sim \mathcal{N}(\mu_i,\,\sigma_i^2)
\end{equation}
encodes uncertainty in gear fault prediction \cite{bnn1}, while physics-based differentiation regularizes the loss function as:
\begin{equation}
    \mathcal{L}_{\text{BNN}} = \mathcal{L}_{\text{data}} 
    + \lambda \, \big\| \nabla_{\theta} f_{\text{phys}} - g_{\text{obs}} \big\|^2,
\end{equation}
where, \( f_{\text{phys}} \) denotes the physics-derived response, and \( g_{\text{obs}} \) the measured behavior. Although effective, this approach requires computationally expensive backpropagation across all layers with limited interpretability. This leads to high training time with lack of comprehensive guideline on optimal sizing of BNNs in line with accuracy guarantees. As shown in Fig. 4(a), the forward and backward pass for optimizing BNN are extensively reliant on data science knowledge and methodologies -- without any physical reflection on the system data.

On the other hand, the \textit{proposed reservoir-inspired method} in Fig. 4(b) restructures this process by transforming gear fault–labeled datasets into empirical histograms conditioned on multiple operating states. Each input vector is mapped through fixed internal dynamics, and only the \textit{readout layer} requires dedicated training efforts. Different from Fig. 4(a), this formulation retains physics fidelity while drastically reducing training complexity. Combined with SHAP explainability-based ranking of important features \cite{shap} and their sensitivity, the proposed method achieves high interpretability, enabling transparent, physics-oriented learning for embedded implementations.

To facilitate adoption of the proposed framework in PE applications, we summarizes key design guidelines derived from the preceding discussions:

\subsection{Model Sizing Based on Physical Structure}
Unlike conventional AI models that require heuristic tuning of network depth and width, the proposed approach enables direct sizing of the model based on system characteristics. As illustrated in Fig. 4(b) , the number of reservoir nodes can be aligned with the number of fault classes and dominant features, given by:
\begin{itemize}
\item The number of output neurons corresponds to the number of fault categories.
\item The number of input connections is determined by the available measurements.
\end{itemize}
This results in a compact architecture with significantly reduced design complexity.

\subsection{Feature Selection via Physical Relevance}
The integration of SHAP-based feature ranking, as shown in Fig. 5 , provides a systematic method for selecting the most informative signals. This allows:
\begin{itemize}
    \item Reduction in sensor count without significant loss in accuracy
\item Prioritization of measurements with high diagnostic value
\end{itemize}

\subsection{Training Efficiency and Computational Benefits}
A major advantage of the proposed framework is that only the readout layer requires training. This leads to reduced computational complexity, faster convergence due to physics-aware initialization and lower memory requirements for embedded implementation. This makes it particularly suitable for real-time diagnostics in industrial drives.

\section{Probabilistic Physics-Aware Learning Theory for Drives}
Before formalizing the probabilistic framework, it is important to interpret the role of feature distributions in the context of motor drive physics. Under each fault condition $\mathrm{k}$, the measured signals do not converge to a single deterministic trajectory but instead exhibit variability due to switching dynamics, load fluctuations, and measurement noise. This variability is not merely stochastic noise; it reflects the underlying physics of the system.

Consequently, the empirical distribution of a feature $\mathrm{x}$ under fault $\mathrm{k}$, denoted as $\mathrm{p(x|k)}$, captures both the dominant system behavior (through its mean) and its uncertainty (through its variance). This statistical paradigm provide an intuitive bridge between physical observations and probabilistic modeling:
\begin{itemize}
    \item \textbf{\textit{Mean}} represents the expected system response under a given fault condition.
\item \textbf{\textit{Variance}} quantifies the sensitivity of that response to operating variations.
\end{itemize}

By embedding the abovementioned variables into the prior distribution of model weights, the proposed approach ensures that each parameter encodes physically meaningful information. This contrasts with conventional neural networks, where weights are initialized randomly and acquire meaning only after extensive training.

For comparative evaluation of computational resources, an uncertainty-aware framework using Bayesian Neural Networks (BNNs) in \cite{bnn1} for gear fault diagnosis is considered. 
%Basically, probabilistic learning is centered around \textit{maximizing the likelihood of events} in each BNN layer using Bayes' theorem:
%\begin{eqnarray}
%    {\underbrace{p(\omega | {X}, {Y})}_{\text{Posterior}} = \frac{ \overbrace{p ({Y}|{X}, \omega)}^{\textbf{Likelihood}}\overbrace{p(\omega)}^{\text{Prior}}}{\underbrace{p({Y}|{X})}_{\text{Evidence}}}}
%\end{eqnarray}
%where, $\mathrm{\omega}$ denote a feature, and $\mathrm{D_{tr}}$ = \{$\mathrm{X}$, $\mathrm{Y}$\} = \{($\mathrm{x}_i$, $\mathrm{y}_i$)\}$_{i=1}^\mathrm{N}$ denote a training dataset with inputs $\mathrm{x}_i \in \mathbb{R}^D$ with their corresponding output, $\mathrm{y}_i \in \{1,...,\mathrm{C}\}$, where $\mathrm{C}$ represents the number of seen gear fault classes in Fig. 2(a). Prior beliefs influence posterior distribution in (1). 

%In this paper, structured and labeled gear fault data is introduced as \textit{empirical histograms} to be treated as natural probabilistic weights \cite{reserv,reserv1} for the first time ever. The input data is processed by being passed through our proposed network structure of neurons, which is designed using probabilistic intrinsic data-turned weights collected under different conditions in Fig. 2. The key idea here is that the AI model’s complexity and nonlinearity are harnessed without needing to train all of its parameters--only the output readout layer is trained. 
Labeled fault datasets collected under distinct operating conditions inherently define condition-dependent feature distributions corresponding to each fault. These empirical distributions can serve as probabilistic priors for model weights, allowing the network parameters to embody the physics and variability of the underlying system behind faulty signatures in gears. This transforms purely random initializations into physically meaningful, probabilistic representations that improve interpretability.

%With a brief theoretical foundation of leveraging probabilistic estimates for quantifying uncertainties, we hypothesize that the state-space relationships between the inputs and output variables in each PE are an unstructured computational space, which needs extrinsic computational efforts in ordering and causation. On top, data-driven organization principles due to lack of data science know-how for PE design engineers can be sub-optimal, that affects its performance. Hence, the structuring of the proposed reservoir model is carried out using SHAP explanations \cite{}, which determines the most important features and their rank that contributes to the AI predictions. As shown in Fig. 2(a), the rank determination and fault conditions are structured into the rows and columns of the proposed reservoir network model.

Let each operating condition or fault type $k$ generate a labeled dataset
\begin{equation}
    \mathcal{D}_k = \{ (x_i,\, y_i=k) \}_{i=1}^{n_k}
\end{equation}
where, $x_i$ denotes a feature vector extracted from measured intrinsic data in Table I.
The empirical histogram of these features under fault $k$ provides an estimate of the conditional probability density:
\begin{equation}
    \hat{p}(x \mid k)
    = \frac{1}{n_k} \sum_{i=1}^{n_k} \delta(x - x_i)
    \approx h_k(x)
\end{equation}
where, $h_k(x)$ represents either a normalized histogram or its smoothed kernel density estimate.

\subsubsection{Mapping the Dataset Histogram to a Probabilistic Weight Prior}
The histogram $h_k(x)$ can be interpreted as a \emph{conditional prior} over the connection strengths $w$ that link each feature component to the output neuron representing fault class $k$. 
A simple and effective mapping is obtained through moment matching between the measured feature distribution and the prior parameters:
\begin{align}
    \mu_k &= \mathbb{E}_{h_k}[x], \\
    \sigma_k^2 &= \mathrm{Var}_{h_k}[x].
\end{align}
These moments define a Gaussian prior for each weight:
\begin{equation}
    \boxed{
    p(w \mid k) = \mathcal{N}\!\left(
        w;\, \mu_k,\, \sigma_k^2 + \tau^2
    \right)}
    \label{eq:gaussian_prior}
\end{equation}
where, $\tau$ is a small regularization constant that accounts for measurement noise or modeling error. 
Each weight's mean and variance therefore inherit the statistical structure of the measured fault responses.
%\subsubsection{Integration into Bayesian Learning}
%During probabilistic training of the readout or classifier layer, the prior in \eqref{eq:gaussian_prior} is combined with the likelihood of observed labels:
%
%\begin{equation}
%    p(w \mid \mathcal{D})
%    \propto
%    \Bigg[
%        \prod_{i=1}^{N}
%        p(y_i \mid x_i, w)
%    \Bigg]
%   \prod_{k} p(w \mid k).
%    \label{eq:posterior}
%\end{equation}
%
%The posterior distribution $p(w \mid \mathcal{D})$ can be approximated using variational inference techniques such as Bayes-by-Backprop, which minimize the Kullback–Leibler (KL) divergence between an approximate posterior $q(w)$ and the true posterior in \eqref{eq:posterior}.
\subsection{SHAP-Guided Placement of Probabilistic Weight Layers}
Following the physics-aware initialization of probabilistic weights, SHAP (SHapley Additive exPlanations) analysis was used to rank the influence of individual reservoir nodes on the final fault classification output. Each SHAP value represents the average contribution of a node’s activation to the model decision, providing a transparent measure of physical relevance.
As shown in Fig. 5, \texttt{Torque} followed by the accelerometers suggests the most important features consistently for each faults. These finalized ranks are then used as a structuring mechanism of the hidden layer placement and dimension as shown in Fig. 6, where the proposed reservoir model transforms input signals into a rich set of temporal features. The only trainable readout layer, introduced in Fig. 6, serves two main purposes:
\begin{itemize}
    \item To map the dynamic, high-dimensional reservoir states (i.e., activations of the internal reservoir nodes) to the target output.
\item To select and combine the most relevant features for final classification.
\end{itemize}
\begin{figure}[t]
    \centering
    \includegraphics[scale=3.65]{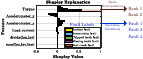}
    \caption{SHAP explanations to rank the important features for gear fault diagnosis--As torque ranks highest in terms of output sensitivity, it features first in reservoir design.}
    \label{Fig_3}
\end{figure}
\begin{figure}[t]
    \centering
    \includegraphics[scale=2.7]{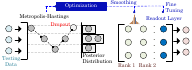}
    \caption{Learning of the readout layer only--this reduces the computational cost, time significantly, and enhances interpretability.}
    \label{Fig_3}
\end{figure}
\subsubsection{Bayesian Readout Training via Bayes-by-Backprop}
\begin{figure*}[h]
    \centering
    \includegraphics[scale=0.95]{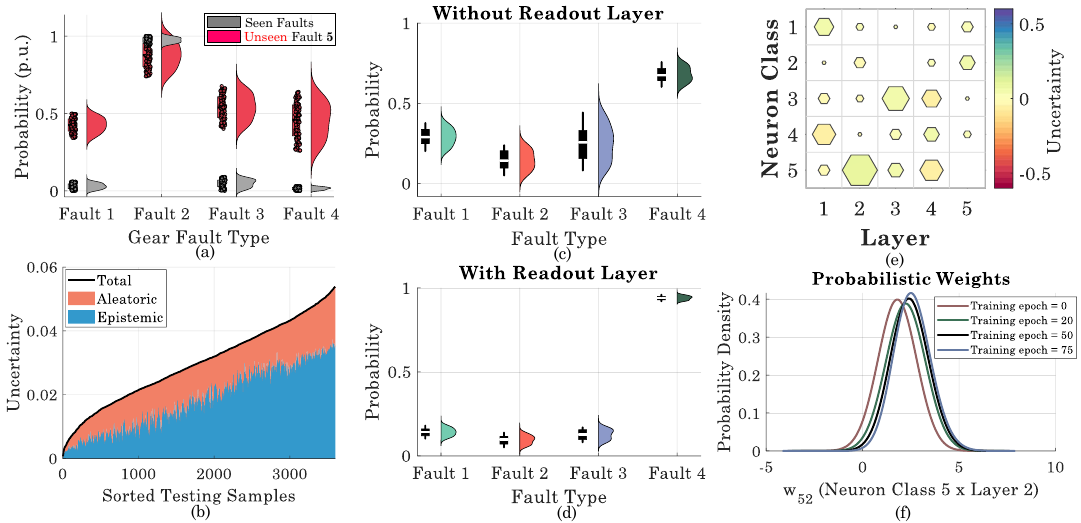}
    \caption{Performance of the proposed reservoir trained on datasets labeled from Fault 1-4 in Table I: (a) Classification accuracy for seen as compared to unseen fault 5, (b) uncertainty plot, (c) without readout layer, (d) with readout layer, (e) interpretability, (f) weight smoothing after training epochs.}
    \label{Fig_32}
\end{figure*}
After ranking the reservoir nodes using SHAP values, the probabilistic weight layer was concentrated on the most informative nodes identified in Fig.~4. 
The subsequent readout layer was trained using the Bayes-by-Backprop (BBB) algorithm \cite{bbb}, which refines the model's probabilistic representation without altering the fixed or smoothly regularized weights in earlier layers. 
\begin{figure}[t]
    \centering
    \includegraphics[scale=1.27]{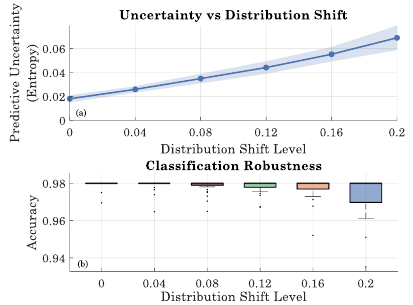}
    \caption{(a) Predictive uncertainty and (b) Classification accuracy under controlled distribution mismatch. While classification accuracy remains relatively stable, the predictive uncertainty increases with distribution shift, indicating reliable detection of out-of-distribution conditions.}
    \label{Fig_3}
\end{figure}
%In this approach, each readout weight is modeled as a Gaussian random variable:
%
%\begin{equation}
 %   w_j \sim \mathcal{N}(\mu_j, \sigma_j^2)
%\end{equation}
%
%where, $(\mu_j, \sigma_j)$ are optimized to approximate the posterior distribution of the true weights given the dataset $\mathcal{D}$. 
Training minimizes the variational free energy (evidence lower bound),
\begin{equation}
    \mathcal{L}_{\text{BBB}}
    =
    \mathbb{E}_{q(w)}[-\log p(y\mid x,w)]
    + \beta\,\mathrm{KL}\!\left[q(w)\,\|\,p(w)\right]
    \label{eq:bbb_loss}
\end{equation}
where, $q(w)=\mathcal{N}(\mu,\sigma^2)$ is the approximate posterior, $p(w)$ is the physics-aware prior defined in~(\ref{eq:gaussian_prior}), and $\beta$ controls the relative weight of the regularization term.
The key idea is to make the internal dynamics rich enough to separate and transform the input signals into linearly separable patterns, which the readout layer can easily exploit.
Since the probabilistic layer already constrains the lower-level weights, BBB primarily smoothens their influence, yielding a compact and interpretable probabilistic readout that condenses the drive’s dynamic behavior into a calibrated fault likelihood. Dropout is implemented via the widely used Metropolis-Hastings algorithm \cite{mena} to approximate posterior distributions. This stage acts as a rapid evaluation of the maximum likelihood for each hidden neuron in Fig. 6.

\section{Performance Evaluation}
%More details on theoretical significance of the causal properties inside the internal reservoir states will be discussed in the full paper.

As shown in Fig. 7, the reservoir model designed using fault data labeled \{1,2,3,4\} (Case 1) is tested on Fault 5 (\textit{unseen} data), providing a representative test of model robustness under unseen conditions. In Fig. 7(a), the predicted probability distributions for the seen gear faults are sharply peaked around their true classes, indicating confident and well-calibrated recognition of Fault 2 during testing.
In contrast, when the unseen fault (Fault~5) is tested, it produces a broader and lower probability response, reflecting the model’s awareness of uncertainty. This confirms that the physics-aware probabilistic initialization and Bayesian readout prevent overconfident predictions on untrained/unseen conditions. Furthermore, Fig. 7(b) decomposes the total predictive uncertainty into aleatoric and epistemic components.
The clear rise in epistemic uncertainty for out-of-distribution samples demonstrates that the reservoir effectively encodes the unseen Fault 5.

Fig.~7(c) and (d) compare the classification performance of the proposed reservoir model without and with the Bayesian readout layer for the four seen fault cases. Without the readout layer, the probability distributions are dispersed across multiple fault classes, showing overlapping confidence and poor class separation. This indicates that the unrefined reservoir activations capture the temporal dynamics but lack a condensed probabilistic representation.

When the Bayesian readout layer is introduced in Fig. 7(d), the class probabilities become sharply peaked around their respective fault types, with minimal overlap between distributions. This demonstrates that the readout effectively condenses the reservoir’s rich dynamics into well-calibrated fault likelihoods, improving both accuracy and interpretability.

Fig.~6(e) visualizes the uncertainty distribution of the probabilistic weights across neuron classes and layers.
The marker size denotes the relative magnitude of each weight, while the color scale represents its associated uncertainty. This layer-wise map allows design engineers to interpret which connections (and fault data) dominate the decision process and how uncertainty is distributed across the customized model.

Fig.~7(f) tracks the evolution of a representative probabilistic weight $w_{52}$ during training.
The progressive narrowing of its probability density for the highest deviating item in Fig. 7(e) demonstrates minimal weight smoothening efforts achieved through Bayesian optimization. As training proceeds, the variance decreases while the mean stabilizes, signifying convergence toward a physically meaningful and well-calibrated weight.

The results in Fig. 8(b) illustrate the behavior of the proposed probabilistic framework under controlled distribution mismatch between training and testing data. As the distribution shift level increases, the classification accuracy exhibits only a gradual decline, indicating that the model retains robust fault discrimination capabilities despite moderate variations in operating conditions. More importantly, the predictive uncertainty in Fig. 8(a) shows a consistent upward trend with increasing mismatch. This behavior reflects the model’s ability to recognize deviations from learned distributions, thereby avoiding overconfident predictions under unseen or out-of-distribution scenarios. Hence, it can be concluded that the proposed approach not only generalizes well but also provides a reliable measure of confidence aligned with the underlying physical variability in the system.
\begin{table}[t]
\centering
\caption{Qualitative Comparison of AI-Based Diagnosis for Power Electronics}
\label{tab:comparison}
\begin{tabular}{lccc}
\hline
\textbf{Method} & {Training Effort} & {Interpretability} & {PE Alignment} \\
\hline
CNN/LSTM   & ++  & --  & -- \\
BNN        & ++  & +   & +  \\
\textbf{Proposed}   & --  & ++  & ++ \\
\hline
\end{tabular}
\vspace{2mm}

\footnotesize{Notation: ++ (High), + (Moderate), -- (Low)}
\end{table}

The comparative summary in Table II highlights conventional deep learning approaches such as CNNs and LSTMs rely on extensive training and large-scale parameter optimization, which often results in limited interpretability and weak alignment with the underlying physical system. While Bayesian Neural Networks (BNNs) introduce uncertainty quantification, they still depend on computationally intensive backpropagation across multiple layers \cite{bnn1} and require careful tuning of model architecture. In contrast, the proposed physics-aware reservoir framework significantly reduces training effort by embedding system knowledge directly into probabilistic weight initialization derived from empirical feature distributions. This results in a compact model where only the readout layer is optimized, leading to lower computational cost and faster deployment. More importantly, the direct linkage between statistical features and physical behavior enables high interpretability, allowing engineers to trace model decisions back to measurable system dynamics and achieving a stronger alignment with power electronic principles.
\section{Conclusion}
This paper introduced a physics-aware probabilistic reservoir modeling framework for motor drive fault diagnosis, where labeled datasets are transformed into statistically grounded model parameters. By mapping empirical feature distributions into probabilistic weights, the proposed method eliminates the need for extensive training while preserving interpretability.

The key contributions of this work can be summarized as follows:
\begin{itemize}
\item A novel mechanism to convert power electronic system data into probabilistic AI parameters
\item A structured reservoir-inspired architecture with minimal training requirements
\item Integration of SHAP-based feature ranking for physically interpretable model design
\item Demonstration of improved uncertainty awareness and robustness under unseen fault conditions
\end{itemize}
The results confirm that the proposed approach not only reduces computational complexity but also provides transparent insights into the relationship between system dynamics and fault signatures. Future work will focus on extending this framework to a broader class of power electronic applications.

\section*{Acknowledgment}

This work is supported by Innovation Fund Denmark under the
Project of Artificial Intelligence for Next-Generation Power Electronics (AI-Power). 

%\section*{SUBMISSION OF THE EXTENDED SUMMARY FOR IPEC}
%\textcolor{red}{An extended summary of Regular Session(RS), Organized Session(OS) and Industry Technology Session(ITS) describing work not previously published or presented must be electronically submitted in a PDF file through the conference website no later than October 30th, 2025.} The submitted extended summary will be reviewed via a peer review process in order to ensure the highest technical quality of the conference. The extended summary should clearly define the salient concepts and novel features of the work. Be sure to mention past or previous works to distinguish your originality from them. The extended summary should be up to 4 pages except Reference, and detailed instructions will be shown on the IPEC-Nagasaki official website, \url{  https://www.ipec2026.org}. Authors will receive notification of acceptance by e-mail on or before January 30th, 2026. Proceedings will be published on IEEE Xplore. Detailed instructions are available on the IPEC-Nagasaki 2026 website http://www.ipec2026.org.

\vspace{12pt}

\end{document}